\documentclass[aps,prd,epsf,nopacs,amsmath,amssymb,graphics,floatfix,onecolumn,11pt]{revtex4}
\usepackage{graphicx}
\usepackage{dcolumn}
\usepackage{bm}
\usepackage{color}
\usepackage{amsmath}
\usepackage{array}
\usepackage{ulem}

\newcommand{\be}{\begin{equation}}
\newcommand{\ee}{\end{equation}}
\newcommand{\ba}{\begin{eqnarray}}
\newcommand{\ea}{\end{eqnarray}}
\newcommand{\ban}{\begin{eqnarray*}}
\newcommand{\ean}{\end{eqnarray*}}

\begin{document}
\title{Extremal Kerr white holes as a source of ultra high energy particles}

\author{$^1$Mandar Patil\footnote{Electronic address: mandar@iitdh.ac.in},
$^2$Tomohiro Harada\footnote{Electronic address: harada@rikkyo.ac.jp}
}

\affiliation{ $^1$ Indian Institute of Technology Dharwad, Dharwad, Karnataka 580011 India. \\ $^{2}$Department of Physics, Rikkyo University,
Toshima-ku, Tokyo 171-8501 Japan.}

\begin{abstract}

We consider a process where two identical massive particles fall inwards, starting from rest
at infinity towards the extremal Kerr black hole, collide outside the event horizon in its vicinity and produce two massless particles.
The center of mass energy of collision between the two particles diverges if one of the particle admits
a specific critical value of the
angular momentum and if the collision takes place at a location arbitrarily close to the event horizon.
Assuming the isotropic emission of particles in the center of mass frame we show that one of the massless
particles produced has divergent conserved energy comparable to the center of mass energy with probability slightly less than half.
This particle enters the black hole event horizon which coincides with the Cauchy horizon, turns back,
and emerges through the white hole event horizon into another asymptotic region in the maximal extension of Kerr spacetime.
Since conserved energy is preserved in this process, it is perceived as a particle with divergent energy by the observer when it reaches infinity.
Thus the extremal white hole appears to be a source of ultra high energy particles.
Similar processes wherein collision takes place slightly inside the black hole event horizon or just outside or inside of
white hole event horizon also produce high energy particles.
\end{abstract}
\maketitle


\section{Introduction}
Earth is bombarded with ultra high energy particles consisting of cosmic rays which are primarily protons and neutrinos \cite{Lisley},\cite{Aartsen}.
The models trying to explain their origin make use of electromagnetic interaction in the essential way, namely Gamma-ray-bursts and
pulsars. In the Gamma-ray-bursts charged particles move back and forth due to the magnetic field and gain energy each time they
cross the shock wave. This process is referred to as Fermi acceleration. In pulsars charged particles are accelerated due to the
electric field generated because of time-varying magnetic field (\cite{Hillas},\cite{Kotera},\cite{Stanev} and references therein). This is referred to as unipolar induction. The high energy neutrinos are believed
to be produced in the process where cosmic rays with energy beyond certain threshold interact with Cosmic-microwave-background \cite{Greisen},\cite{Zatsepin}.

We would like to make an attempt to come up with models which make use of gravity, rather than electromagnetic
interaction to produce ultra high energy particles. For this purpose, the collisional Penrose process wherein one can extract energy coming from rotating spacetime
admitting so called ergo-region is useful \cite{Penrose},\cite{Piran1},\cite{Piran2}. It is a process where two particles fall inwards and collide inside the ergo-region
to produce two particles. One of the particles produced in the collision can be launched onto the trajectory with negative conserved
energy, while other particle escapes to infinity. From the conservation of conserved energy in the collision process it follows that
the particle escaping to infinity carries energy larger than the initial energy of the two colliding particles leading to the extraction
of rotational energy from the spacetime.

Authors of this paper along with others had studied the collisional Penrose process in the over-spinning rotating Kerr spacetime
describing the exterior of ultra compact Kerr super-spinor whose size smaller than the gravitational radius \cite{Patil}. We considered
a head-on collision between the radially ingoing particle
and an initially ingoing particle which turns back due to the angular momentum barrier and is the radially outgoing collision at the moment
of collision. The center of mass energy of the collision diverges if the collision takes place at a specific location
if the spin parameter is arbitrarily close to the extremal value from above. Assuming an isotropic distribution of two massless particles
produced in the collision in the center of mass frame, we show that one of the two massless particle produced in the collision has divergent conserved energy
with probability slightly less than half. Irrespective of whether this particle is moving in radially inward or outward direction, it inevitably
escapes to infinity. It is perceived as the particle with divergent energy by an observer at infinity. Thus near-extremal over-spinning
Kerr geometry acts as a source of ultra high energy particles. Large amount of rotational energy
is extracted from the Kerr spacetime by collisional Penrose process which leads to the production of ultra high energy particles.
This process was generalized to spacetime satisfying certain properties \cite{Zas}. Wormholes can also be used to generate ultra high energy
particles \cite{Bambi}.

In this paper we present a process of formation of ultra high energy particles in the maximally rotating Kerr spacetime. The maximal
extension of extremal Kerr spacetime consists of infinitely many asymptotically flat regions of spacetime each with black hole and
white hole event horizons \cite{BL},\cite{Carter}. Interior of the black hole event horizon in one asymptotic region corresponds to the interior of
white hole event horizon with respect to the next asymptotic region. We consider a process of collision of two identical massive
particles, which start at rest at time-like past infinity in one of the asymptotic region and fall towards the black hole event horizon.
If the conserved angular momentum of one of the particles is set to a specific critical value and the collision takes place at the
location arbitrarily close to the event horizon, the center of mass energy of collision displays a divergence \cite{BSW},\cite{Harada}. We assume that
two massless particles are produced in the collision. Assuming that they are produced isotropically in the center of mass frame
we show that one of the particles has conserved energy which is comparable to the center of mass energy and thus shows divergence,
with the probability slightly less than half. This particle enters the black hole event horizon which coincides with the
Cauchy horizon, turns back and emerges from the white hole horizon in the next asymptotic region. In this process conserved energy is preserved.
Thus when it reaches infinity it is perceived as the particle with divergent energy by the observer at infinity. Thus extremal
white hole can serve as a source of ultra high energy particles. On the other hand if massless particle escapes to infinity
in the same asymptotic region as that of collision it has energy almost one order of magnitude greater than
of colliding particles \cite{Bejger},\cite{Harada1},\cite{Schnittman}.

In section II we describe Kerr spacetime and its features. In section III we discuss geodesic motion of massive and massless particles
in Kerr spacetime. In section IV we describe the process of ultra high energy collisions near the event horizon of extremal
Kerr black hole. In section V we discuss the process of generation of high energy particles and its connection with white hole.
In section VI we end the discussion with concluding remarks.

\section{Extremal Kerr black hole metric}
In this section we discuss the metric of extremal Kerr spacetime and describe its properties.
We write down the Kerr metric in Boyer Lindquist coordinate system as
\begin{equation}
ds^2 = -\left(1-\frac{2Mr}{\Sigma}\right)dt^2+\left(r^2+a^2+\frac{2Mra^2}{\Sigma} \sin^2 \theta\right) \sin^2 \theta d\phi^2
-\frac{2Mra}{\Sigma} \sin^2\theta dtd\phi+ \frac{\Sigma}{\Delta}dr^2+ \Sigma d\theta^2 \ ,
\label{Kerrmetric}
\end{equation}
where the quantities $\Delta$ and $\Sigma$ appearing in the expression above are given by
$\Delta=\left(r^2-2Mr+a^2\right)$ and $\Sigma=\left(r^2+a^2\cos^2\theta\right)$. There are two parameters in the Kerr metric,
mass $M$ and spin $a$. The Kerr spacetime admits event horizon and Cauchy horizon when $a \le M$ and they are located at
$r=M\pm \sqrt{M^2-a^2}$. The region inside the event horizon is referred to as black hole. The Kerr spacetime admits two Killing vectors $\partial_t$ and $\partial_{\phi}$
that are associated with time translation and axial symmetry. When $a=M$, i.e., when the spin parameter admits
maximum possible value for the Kerr metric to represent Kerr black hole, the black hole is known as extremal Kerr black hole.
In this case the event horizon and Cauchy horizon coincide and occur at $r=M$. The metric both outside and inside the extremal Kerr
black hole is given by
\begin{equation}
ds^2 = -\left(1-\frac{2Mr}{\Sigma}\right)dt^2+\left(r^2+M^2+\frac{2M^3r}{\Sigma} \sin^2 \theta\right) \sin^2 \theta d\phi^2
-\frac{2M^2r}{\Sigma} \sin^2\theta dtd\phi+ \frac{\Sigma}{\Delta}dr^2+ \Sigma d\theta^2 \ ,
\label{Kerrmetric1}
\end{equation}
where $\Delta=(r-M)^2$ and $\Sigma=\left(r^2+M^2\cos^2\theta\right)$. The Boyer Lindquist coordinate system breaks down at the location of event horizon.
In order to smoothly connect the region inside and outside the event horizon we use ingoing or outgoing null coordinates in which
metric is well-behaved at the horizon.

We introduce ingoing null coordinate system $(v,r,\theta,\varphi)$ which is related to the Boyer Lindquist coordinate system by
\begin{equation}
 dv=dt+(r^2+M^2)\frac{dr}{\Delta} ~~,~~ d\varphi=d\phi+M \frac{dr}{\Delta}.
 \label{ct1}
\end{equation}
In this coordinate system Kerr metric is given by
\begin{eqnarray}
 ds^2=&&-\left(1-\frac{2Mr}{\Sigma}\right)dv^2+2dvdr+\Sigma d\theta^2+\frac{\left( (r^2+M^2)^2-M^2\Delta \sin^2\theta  \right)
 \sin^2\theta}{\Sigma}\sin^2\theta d\varphi^2 \nonumber \\
 &&- 2 M \sin^2\theta d\varphi dr -\frac{4M^2r}{\Sigma} \sin^2\theta d\varphi dv.
 \label{inc}
\end{eqnarray}
It admits Killing vectors $\partial_{v}$ and $\partial_{\varphi}$.

Similarly the metric can also be written in outgoing null coordinates $(u,r,\theta,\varphi)$ related to the Boyer Lindquist coordinates by relation
\begin{equation}
 du=dt-(r^2+M^2)\frac{dr}{\Delta} ~~,~~ d\varphi=d\phi- M \frac{dr}{\Delta}.
 \label{ct2}
\end{equation}
In this coordinate system Kerr metric is given by
\begin{eqnarray}
 ds^2=&&-\left(1-\frac{2Mr}{\Sigma}\right)du^2-2dudr+\Sigma d\theta^2+\frac{\left( (r^2+M^2)^2-M^2\Delta \sin^2\theta  \right)
 \sin^2\theta}{\Sigma}\sin^2\theta d\varphi^2 \nonumber \\
 &&+ 2 M \sin^2\theta d\varphi dr -\frac{4M^2r}{\Sigma} \sin^2\theta d\varphi du.
 \label{outc}
\end{eqnarray}
The Killing vectors associated with this metric is $\partial_{u}$ and $\partial_{\varphi}$.

\begin{figure}
\begin{center}
\includegraphics[width=0.5\textwidth]{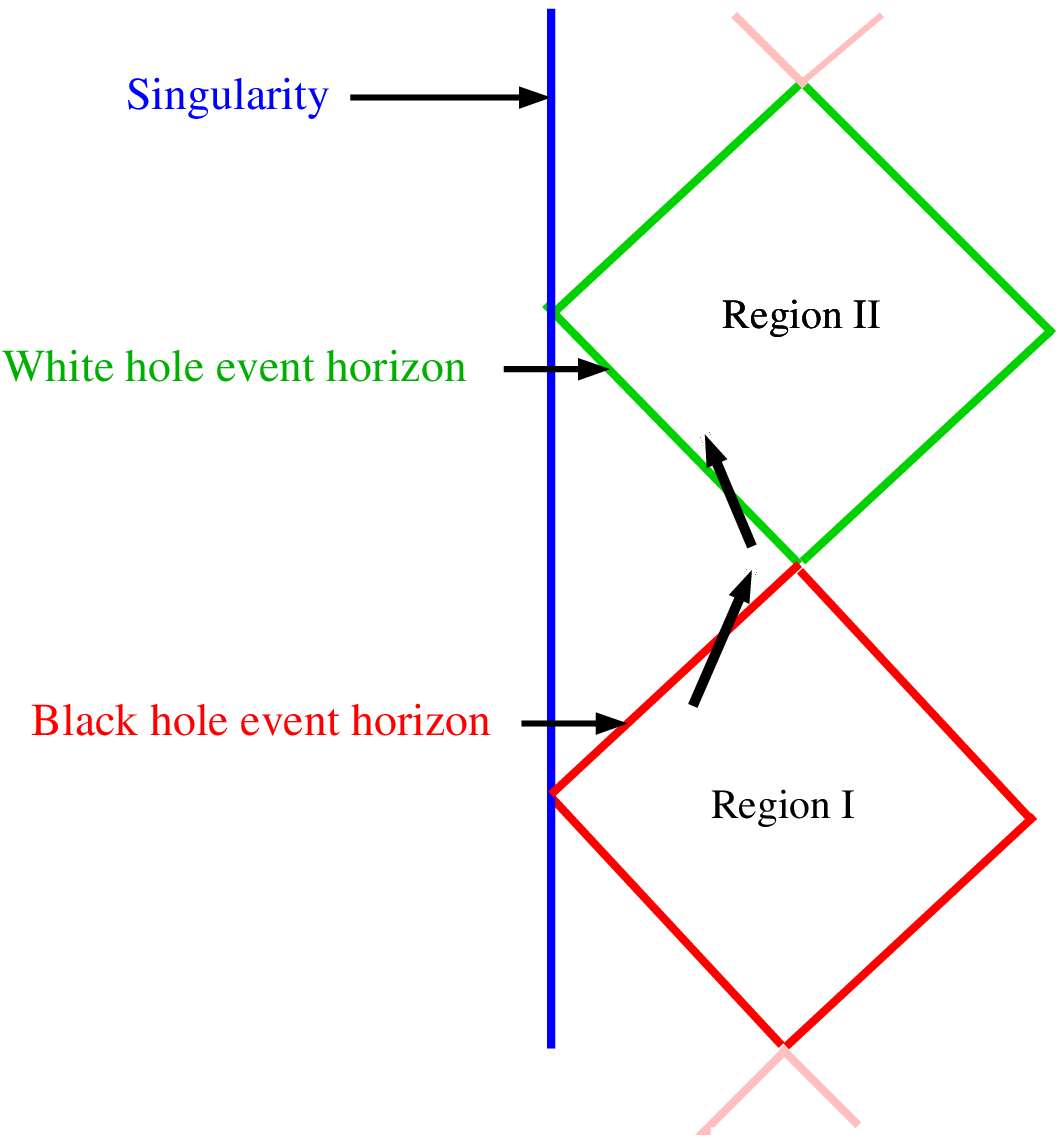}
\caption{ Penrose diagram of maximally extended extremal Kerr spacetime is depicted in the figure. Region I and Region II
are two asymptotic regions under consideration. The time-like singularity, black hole horizon and white hole horizons are depicted
in the figure. The massless particle enters the black hole event horizon in Region I, turns back and emerges out of
white hole event horizon in Region II. Its motion is depicted by thick black arrows.}
\label{penrose}
\end{center}
\end{figure}

The Penrose diagram of the maximally extended extremal Kerr spacetime is shown in Fig.\ref{penrose}. It contains a infinite
sequence of asymptotically flat regions. Each of such asymptotic region admits a black hole and white hole event horizons.
The black hole region for a given asymptotic region corresponds to the white hole for the next asymptotic region.
One can enter the black hole event horizon in one asymptotically flat region
and emerge into the upper asymptotically flat region from the white hole event horizon.
As we shall see later we are interested in the situation where massless particle with large conserved energy produced in the collision just
outside, enters the black hole event horizon, admits a turning point (this is possible since
we are dealing with the region inside Cauchy horizon) and emerges out of a white hole event horizon into next asymptotically
flat region of spacetime.

\section{Geodesics of massless and massive particles }
In this section we describe the geodesic motion of massive as well as massless particles in the Kerr spacetime.

\subsection{Massive particle geodesic motion}

We consider a massive particle following geodesic motion on the equatorial plane $\theta=\frac{\pi}{2}$ outside
the event horizon. The four-velocity $U$ of such a particle is given by
an expression
\begin{eqnarray}
 U^{t} &=& \frac{1}{(r-M)^2}\left(E\left(r^2+M^2+\frac{2M^3}{r}\right)-\frac{2M^2}{r}L\right) \ , \nonumber \\ \nonumber \\
 U^{\phi} &=& \frac{1}{(r-M)^2} \left( \frac{2M^2}{r}E+\left(1-\frac{2M}{r}\right)L \right) \ , \nonumber \\ \nonumber \\
 U^{r} &=& \sigma \sqrt{E^2-1+\frac{2M}{r}-\frac{L^2-M^2\left(E^2-1\right)}{r^2}+\frac{2M \left(L-ME\right)^2}{r^3}} \ , \nonumber \\ \nonumber \\
 U^{\theta} &=& 0 \ ,
 \label{velmav}
\end{eqnarray}
where $E=-\partial_{t}\cdot U$ and $L=\partial_{\phi}\cdot U$ are the conserved energy and angular momentum per unit mass respectively and
$\sigma=\pm 1$ depending on whether the particle moves in radially outwards or inwards. These equations are obtained by solving the
equations $E=-\partial_{t}\cdot U$ and $L=\partial_{\phi}\cdot U$ which define the conserved quantities associated with time translation
and azimuthal symmetries, along with the normalization condition $U \cdot U=-1$ for the four-velocity of massive particle.

We consider a case where $E=1$. From Eq.\ref{velmav} we can infer that it corresponds to the particle that is at rest at infinity,
if at all it starts out or ends up there.
The expression for the four-velocity now can be obtained by setting $E=1$ in Eq.\ref{velmav} and is given by
\begin{eqnarray}
 U^{t} &=& \frac{1}{(r-M)^2}\left(\left(r^2+M^2+\frac{2M^3}{r}\right)-\frac{2M^2}{r}L\right) \ , \nonumber \\ \nonumber \\
 U^{\phi} &=& \frac{1}{(r-M)^2} \left( \frac{2M^2}{r}+\left(1-\frac{2M}{r}\right)L \right) \ , \nonumber \\ \nonumber \\
 U^{r} &=& \sigma \sqrt{\frac{2M}{r}-\frac{L^2}{r^2}+\frac{2M \left(L-M\right)^2}{r^3}} \ , \nonumber \\ \nonumber \\
 U^{\theta} &=& 0 \ ,
 \label{velmav1}
\end{eqnarray}
The particle admits a turning point when $U^{r}=0$. From Eq.\ref{velmav1} the radial coordinate of turning point is given by
\begin{equation}
 r=r_{\pm}=\frac{L^2}{4M}\left( 1 \pm \frac{16M^2(L-M)^2}{L^4} \right).
 \label{tp}
\end{equation}
The turning point would exist only for those values of $L$ for which the quantity under the square root sign is positive. For this
we must have
\begin{equation}
  -2(1+\sqrt{2})M <  L < 2M.
 \label{ltp}
\end{equation}
We are interested in the situation where particle falls inwards starting from rest at infinity. This implies that if $L<-2(1+\sqrt{2})M $,
it will turn back and starts moving in outward direction at $r=r_{+}>(1+\sqrt{2})^2M$ and
if $L>2M$ it will turn back at location $r>M$. If $L=2M$, which we refer to as critical angular momentum,
the particle will approach event horizon located at $r=M$ asymptotically. As we shall see this is the case of interest to us.

\subsection{Massless particle geodesic motion}
We consider the motion of massless particle on the equatorial plane of Kerr spacetime. We shall discuss the motion both outside and
inside the black hole. The four-velocity of the massless particle is given by
\begin{eqnarray}
 U^{t} &=& \frac{1}{(r-M)^2}\left(\left(r^2+M^2+\frac{2M^3}{r}\right)-\frac{2M^2b}{r}\right) \ , \nonumber \\ \nonumber \\
 U^{\phi} &=& \frac{1}{(r-M)^2} \left( \frac{2M^2}{r}+\left(1-\frac{2M}{r}\right)b \right) \ , \nonumber \\ \nonumber 
 U^{r} &=& \sigma\sqrt{1-\frac{\left(b^2-a^2\right)}{r^2}+\frac{2M\left(b-a\right)^2}{r^3}} \nonumber \ , \\ \nonumber \\
 U^{\theta} &=& 0 \ ,
 \label{velmal}
\end{eqnarray}
where $b=\frac{L}{E}$ is the impact parameter. $E$ and $L$ are conserved energy and angular momenta (not per unit mass as in the case
of massive particle) respectively. The equations of motion above can be derived using the equations $E=-\partial_{t}\cdot U$ and
$L=\partial_{\phi}\cdot U$ augmented with the normalization condition $U \cdot U=0$ for the massless particle. We further impose the
new parametrization $\lambda \to E\lambda$.

\begin{figure}
\begin{center}
\includegraphics[width=0.6\textwidth]{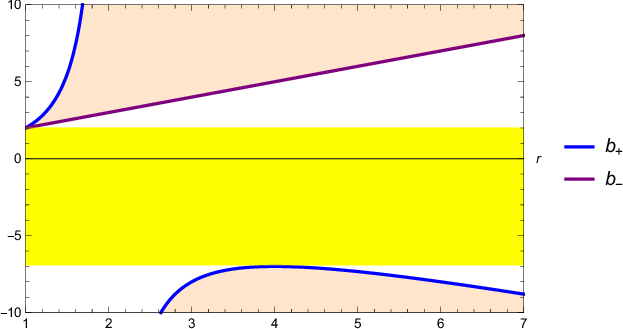}
\caption{ We plot $b_{+}$ and $b_{-}$ as a function of $r$ in the region outside event horizon in the units where $M=1$. $b_{+}$ increases from $2$ to infinity as we go from
$r=1$ to $r=2$. It goes to $-\infty$ at $r=2$ and as $r\to \infty$. It admits a maximum at $r=4$ with the value $-7$.
$b_{-}$ increases from $2$ to $\infty$ linearly as we go from $r=1$ to $\infty$. The prohibited region is depicted in pink,
whereas the yellow region is relevant for particle that fall from infinity and reach event horizon.
}
\label{br}
\end{center}
\end{figure}

We first focus our attention on the massless particle that moves outside the event horizon. The particle would
admit a turning point in radial direction if $U^r=0$. From Eq.\ref{velmal}, the impact parameter necessary for the particle to turn back at radial
coordinate $r$ is given by
\begin{equation}
b_{+}(r)=-(r+M)+\frac{4M^2}{(2M-r)}~~,~~b_{-}(r)=(r+M).
\label{bval}
\end{equation}
We consider two cases. The first case is one where $r>2M$. In this case it is easy to verify that both $b_{+}$ is negative
and $b_{-}$ is positive. $b_{-}$ is momotonically increasing function which is $b_{-}=3M$ at $r=2M$ and it goes to infinity as $r \to \infty$.
Whereas $b_{+}(r)$ tends to minus infinity at $r=2M$ as well as $r\to \infty$ and it admits maximum at $r=4M$,
where $b_{+,m}=-7M$. We now consider the second case where $M<r<2M$. Since $b_{+}-b_{-}=2M-2M=0$ at $r=M$
and $\frac{d}{dr}\left(b_{+}-b_{-}\right)>0$ when $M<r<2M$, we have $b_{+}>b_{-}$. While $b_{-}$ increases linearly from $2M$ to $3M$
as we go from $r=M$ to $r=2M$, $b_{+}$ increases from $2M$ to $\infty$. $b_{+}$ and $b_{-}$ are plotted in Fig.\ref{br}.

We are interested in the massless particles generated close to the horizon in the collision event.
So the particles generated at radial coordinate $r$ that are moving in the radially outward direction will escape to inifnity
if $b$ is in the range
\begin{equation}
\sigma=+1~~,~~ -7M  < b < b_{-}(r),
 \label{brange1}
\end{equation}
while the particles moving in the radially inward direction escape as long as
\begin{equation}
\sigma=-1 ~~,~~ 2M  < b < b_{-}(r).
 \label{brange2}
\end{equation}
All other massless particles would enter the horizon.

\begin{figure}
\begin{center}
\includegraphics[width=0.6\textwidth]{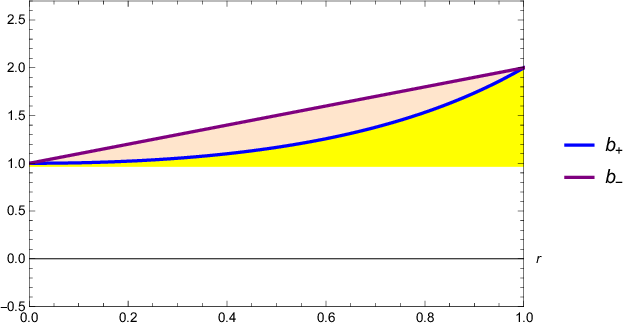}
\caption{ We plot $b_{+}$ and $b_{-}$ as a function of $r$ inside the event or Cauchy horizon in the units where $M=1$. Both $b_{+}$
and $b_{-}$ decrease from $2$ to $1$ as we go from $r=1$ to $r=0$  while $b_{+}<b_{-}$ in this range.
The prohobited region is depicted in pink, while the yellow region is relevent for the massless particles that enter
the horizon, turn back and come out of white hole event horizon to the next asymptotic region.}
\label{br1}
\end{center}
\end{figure}

Now we consider the situation where massless particle moves inside the event or Cauchy horizon. There is a time-like singularity at $r=0$
and spacetime ends there. The particle which enters the event horizon which moves in the inward direction  will turn back when
$U^{r}=0$ and this happens when impact parameter takes values
\begin{equation}
b_{+}(r)=-(r+M)+\frac{4M^2}{(2M-r)}~~,~~b_{-}(r)=(r+M).
\label{bval1}
\end{equation}
Both $b_{+}$ and $b_{-}$ take values $2M$ at $r=M$ and $M$ when $r=0$ while $b_{+}<b_{-}$ in the entire range. We have plotted
$b_{+}$ and $b_{-}$ in Fig.\ref{br1}. It is quite obvious that if impact parameter $b$ is in the range
\begin{equation}
M < b < 2M,
\label{turnb}
\end{equation}
the massless particle would turn back and then it will emerge out of white hole region in the next asymptotic region as shown in the Fig.1.
All other particles hit the singularity.

We are interested in the situation where massless particles are formed close to the event horizon, enter the black hole, turn back and
exit through the white hole horizon and enter the new asymptotically flat region of maximally extended extremal Kerr spacetime.
While Boyer Lindquist coordinates are well defined inside as well as of outside black hole or white hole horizon, they break
down at the location of black hole or white hole horizon. Thus we need to introduce coordinates that are well-defined at the
horizon. Close to the black hole horizon we introduce ingoing null coordinates employed in Eq.\ref{inc} and close to the white hole horizon we introduce
outgoing null coordinates employed in Eq.\ref{outc}.

Description of exterior of black hole is in terms of Boyer Lindquist coordinates. In order to deal with massless particles entering
black hole we introduce ingoing null coordinates by making coordinate transformation in Eq.\ref{ct1}. The constants of motion
associated with the null geodesic motion in the two coordinate systems are related by
\begin{eqnarray}
 e=-\partial_{v}\cdot U = \left(1-\frac{2M}{r}\right)\dot{v}-\dot{r}+\frac{2M^2}{r}\dot{\varphi}
  = \left(1-\frac{2M}{r} \right)\dot{t}+\frac{2M^2}{r}\dot{\phi} =-\partial_{t}\cdot U
  = E ,
\end{eqnarray}
and
\begin{eqnarray}
 l=\partial_{\varphi}\cdot U = \left( r^2+M^2+\frac{2M^3}{r}\right)\dot{\varphi}-M\dot{r}-\frac{2M^2}{r}\dot{v}
  = \left(1-\frac{2M}{r} \right)\dot{t}+\frac{2M^2}{r}\dot{\phi} =\partial_{\phi}\cdot U
  = L
\end{eqnarray}
Thus conserved energy $E$ and angular momentum $L$ associated with Killing vectors $\partial_{t}$
and $\partial_{\phi}$ in Boyer Lindquist coordinates are same as $e$ and $l$, the constants of motion associated with Killing vectors
$\partial_{v}$ and $\partial_{\varphi}$ in ingoing null coordinates.

After we enter the black hole horizon we employ inverse of the transformation Eq.\ref{ct1} and reintroduce the Boyer Lindquist
coordinates. We can easily show that constants of motion associated with null geodesic motion inside the black hole
are same as that of outside, namely $E$ and $L$. Thus the impact parameter $b=\frac{L}{E}$ is also the same.

If the massless particle turns back inside the black hole, which would  happen if the condition given by Eq.\ref{turnb} holds,
it will emerge out of white hole event horizon into the new asymptotic region.
In order to describe this we introduce outgoing null coordinates via transformation Eq.\ref{outc}. We can show that constants
of motion $\partial_{u}$ and $\partial_{\varphi}$ are the same as that of $\partial_{t}$ and $\partial_{\phi}$.
\begin{eqnarray}
 e'=-\partial_{u}\cdot U = \left(1-\frac{2M}{r}\right)\dot{u}+\dot{r}+\frac{2M^2}{r}\dot{\varphi}
  = \left(1-\frac{2M}{r} \right)\dot{t}+\frac{2M^2}{r}\dot{\phi} =-\partial_{t}\cdot U
  = E ,
\end{eqnarray}
and
\begin{eqnarray}
 l'=\partial_{\varphi}\cdot U = \left( r^2+M^2+\frac{2M^3}{r}\right)\dot{\varphi}+M\dot{r}-\frac{2M^2}{r}\dot{v}
  = -\frac{2M^2}{r}\dot{t}+\left(r^2+M^2+\frac{2M^3}{r} \right)   =\partial_{\phi}\cdot U
  = L
\end{eqnarray}
After entering the white hole horizon we can reintroduce Boyer Lindquist coordinates and constants of motion are still old
$E$ and $L$ and consequently the impact parameter is also same. So massless particle emerges out of white hole with exactly the same energy
with which it was generated in the trailing asymptotic region. We shall use this result later.

\section{High energy collisions near event horizon}

We consider two identical particles  with mass $m$ which fall towards the black hole starting from rest at infinity.
So for these particles conserved energies are $E_1=1$ and $E_2=1$. We assume that for one of the particles angular momentum
is given by $L_1=2M$ so that this particle asymptotically approaches the event horizon located at $r=M$. While the other
particle has the angular momentum in the range $-2(1+\sqrt{2})M <  L_2 < 2M$ so that this particle arrives at the horizon with
finite radial component of velocity. Both the particles travel in the inward direction at the location of collision
so that $\sigma_1=\sigma_2=-1$. We assume that the collision occurs at the location which is very close to the event
horizon, i.e., $r=M(1+\epsilon)$ with $\epsilon \to 0$. We first present the calculation for general $~L_1,~L_2$ and $r$ and later on we substitute the
chosen values, while we assume from the outset that $E=1$ and $\sigma=-1$ for massive colliding particles.

\subsection{LNRF}
We first make a transition to the locally non-rotating frame (LNRF) which is the tetrad that executes circular motion around
the black hole with frame-dragging frequency. On the equatorial plane tetrad components are given by
\begin{eqnarray}
e_{\mu}^{(t)} &~:~& e_{t}^{(t)}=\frac{(r-M)}{r^2+M^2+\frac{2M^3}{r}} \ ; e_{\phi}^{(t)}=0 \ ; e_{r}^{(t)}=0 \ ; e_{\theta}^{(t)}=0 \ , \nonumber \\ \nonumber \\
e_{\mu}^{(\phi)} & ~:~& e_{t}^{(\phi)}=-\frac{\frac{2M^2}{r}}{\sqrt{\left(r^2+M^2+\frac{2M^3}{r}\right)}} \ ; e_{\phi}^{(\phi)}=\sqrt{\left(r^2+M^2+\frac{2M^3}{r}\right)} \ ; e_{r}^{(\phi)}=0 \ ; e_{\theta}^{(\phi)}=0 \ , \nonumber \\ \nonumber \\
e_{\mu}^{(r)} & ~:~& e_{t}^{(r)}=0 \ ; e_{\phi}^{(r)}=0 \ ; e_{r}^{(r)}=\frac{r}{(r-M)} \ ; e_{\theta}^{(r)}=0 \ , \nonumber \\ \nonumber \\
e_{\mu}^{(\theta)} & ~:~& e_{t}^{(\theta)}=0 \ ; e_{\phi}^{(\theta)}=0 \ ; e_{r}^{(\theta)}=0 \ ; e_{\theta}^{(\theta)}=\sqrt{\left(r^2+M^2+\frac{2M^3}{r}\right)} \ .
\label{lnrf}
\end{eqnarray}

Components of any vector in the LNRF are related to the components in the Boyer Lindquist coordinates by the relation
\begin{equation}
 V^{(\mu)}_{LNRF}=e_{\nu}^{(\mu)}V^{\nu}.
\end{equation}

Components of four-velocity of massless particles in LNRF can be written as
\begin{eqnarray}
  U^{(t)}_{LNRF} &=& \frac{1}{(r-M)}\frac{\left(\left(r^2+M^2+\frac{2M^3}{r}\right)
  -\frac{2M^2b}{r}\right)}{\sqrt{\left(r^2+M^2+\frac{2M^3}{r}\right)}}  \ , \nonumber \\ \nonumber \\
  U^{(\phi)}_{LNRF} &=& \frac{b}{\sqrt{\left(r^2+M^2+\frac{2M^3}{r}\right)}} \ ,\nonumber \\ \nonumber \\
   U^{(r)}_{LNRF} &=& \sigma \frac{r}{(r-M)} \sqrt{1-\frac{\left(b^2-M^2\right)}{r^2}+\frac{2M\left(b-M\right)^2}{r^3}} \ , \nonumber  \\ \nonumber \\
 U^{(\theta)}_{LNRF} &=&  0 \ .
 \label{lnrfmal}
\end{eqnarray}

We define $A$, $B$ and $C$ as non-vanishing components of sum of velocities of two colliding particles and they are given by
\begin{eqnarray}
 A &=& U^{(t)}_{1,LNRF}+U^{(t)}_{2,LNRF} \nonumber \\ 
   &=&  \frac{1}{(r-M)}\frac{\left(\left(r^2+M^2+\frac{2M^3}{r}\right)
  -\frac{2Ma}{r}L_{1}\right)}{\sqrt{\left(r^2+M^2+\frac{2M^3}{r}\right)}}
  + \frac{1}{(r-M)}\frac{\left(\left(r^2+M^2+\frac{2M^3}{r}\right)
  -\frac{2M^2}{r}L_{2}\right)}{\sqrt{\left(r^2+M^2+\frac{2M^3}{r}\right)}} \ , \nonumber \\ \nonumber \\
 B &=&  U^{(r)}_{1,LNRF}+U^{(r)}_{2,LNRF} \nonumber \\ 
   &=& - \frac{r}{(r-M)} \sqrt{\frac{2M}{r}-\frac{L_{1}^2}{r^2}+\frac{2M \left(L_{1}-M\right)^2}{r^3}} \nonumber \\
         && - \frac{r}{(r-M)} \sqrt{\frac{2M}{r}-\frac{L_{2}^2}{r^2}+\frac{2M \left(L_{2}-M\right)^2}{r^3}} \ , \nonumber \\ \nonumber \\
C &=& U^{(\phi)}_{1,LNRF}+U^{(\phi)}_{2,LNRF}  \nonumber \\ 
  &=& \frac{L_{1}}{\sqrt{\left(r^2+M^2+\frac{2M^3}{r}\right)}} +\frac{L_{2}}{\sqrt{\left(r^2+M^2+\frac{2M^3}{r}\right)}} \ .
\label{ABC}
\end{eqnarray}

When we substitute the values of $L_1$, $L_2$ and $r$, to the sub-leading order in $\epsilon$ we get
\begin{equation}
 A=\frac{(2-l_2)}{\epsilon}+(2+l_2)~~,~~ B=-\frac{(2-l_2)}{\epsilon}-(l_2+\sqrt{2})~~,~~C=\left(1+\frac{l_2}{2} \right),
\end{equation}
where $l_2=\frac{L_2}{M}$.

It is evident from the expression above that net radial component of velocity is negative, meaning that the center of mass
moves in the radially inward direction. Whereas the sign of $C$ determines whether the center of massrotates in the same sense or opposite
sense as that of the rotation associated with spacetime.

\subsection{Center of mass frame}
We now make a transition to the center of mass frame, a tetrad in which the spatial components of net-velocity of two colliding
particles is zero and the time-component yields the center of mass energy of collision. The center of mass frame is related to
LNRF by means of a Lorentz transformation given below. The Lorentz transformation is combination of two components. First one is a
rotation which orients the net-component of spatial velocity in the radial direction, followed by the boost which will kill
spatial component of net-velocity. The Lorentz transformation is given by
\begin{equation}
 V_{cm}^{(\mu)}=\Lambda_{boost~(\nu)}^{(\mu)} \Lambda_{rot~(\sigma)}^{(\nu)}V^{(\sigma)}_{LNRF},
\end{equation}
where
\begin{equation}
 \Lambda_{rot}=\begin{bmatrix}
 1 & 0 & 0 & 0 \\
  0 & \frac{B}{\sqrt{B^2+C^2}} & -\frac{C}{\sqrt{B^2+C^2}} & 0 \\
  0 & \frac{C}{\sqrt{B^2+C^2}}  & \frac{B}{\sqrt{B^2+C^2}} & 0 \\
  0 & 0 & 0 & 1
  \end{bmatrix}
  ~~~,~~~
  \Lambda_{boost}=\begin{bmatrix}
                   \frac{A}{\sqrt{A^2-B^2-C^2}} & 0 & -\frac{\sqrt{B^2+C^2}}{\sqrt{A^2-B^2-C^2}}  & 0 \\
                   0 & 1 & 0 & 0 \\
                   -\frac{\sqrt{B^2+C^2}}{\sqrt{A^2-B^2-C^2}} & 0 & \frac{A}{\sqrt{A^2-B^2-C^2}}  & 0 \\
                   0 & 0 & 0 & 1.
                  \end{bmatrix}
\end{equation}
Transformation directly from Boyer Lindquist coordinates to center of massframe is given by
\begin{equation}
 V_{cm}^{(\mu)}=\Lambda_{boost~(\nu)}^{(\mu)} \Lambda_{rot~(\sigma)}^{(\nu)}  e_{\delta}^{(\sigma)} \ V^{\delta}.
\end{equation}
The center of mass energy of collision shows divergence in the limit where $\epsilon \to 0$, i.e., when collision
happens at the location arbitrarily close to the event horizon of extremal Kerr black hole located at $r=M$.

Using the transformation above we can write down the expression for the four-velocity of the massless particles in the
center of mass frame.
\begin{eqnarray}
 U^{(t)}_{cm} &=& \frac{A}{\sqrt{A^2-B^2-C^2}} \ \frac{1}{(r-M)}\frac{\left(\left(r^2+M^2+\frac{2M^3}{r}\right)-\frac{2M^2b}{r}\right)}{\sqrt{\left(r^2+M^2+\frac{2M^3}{r}\right)}}
 - \frac{C}{\sqrt{A^2-B^2-C^2}} \ \frac{b}{\sqrt{\left(r^2+M^2+\frac{2M^3}{r}\right)}}
 \nonumber \\
 &&- \frac{B}{\sqrt{A^2-B^2-C^2}} \sigma \frac{r}{(r-M)}\sqrt{1-\frac{\left(b^2-M^2\right)}{r^2}+\frac{2M\left(b-M\right)^2}{r^3}} \ ,\nonumber \\
  \nonumber \\
U^{(\phi)}_{cm} &=& -\frac{C}{\sqrt{B^2+C^2}} \ \sigma \frac{r}{(r-M)} \sqrt{1-\frac{\left(b^2-M^2\right)}{r^2}+\frac{2M\left(b-M\right)^2}{r^3}} \nonumber \\
 &&+ \frac{B}{\sqrt{B^2+C^2}} \ \frac{b}{\sqrt{\left(r^2+M^2+\frac{2M^3}{r}\right)}}  \ , \nonumber  \\ 
 U^{(r)}_{cm} &=& -\frac{\sqrt{B^2+C^2}}{\sqrt{A^2-B^2-C^2}} \ \frac{1}{(r-M)}\frac{\left(\left(r^2+M^2+\frac{2M^3}{r}\right)-\frac{2M^2 b}{r}\right)}{\sqrt{\left(r^2+M^2+\frac{2M^3}{r}\right)}} \nonumber \\
  &&+ \frac{AC}{\sqrt{A^2-B^2-C^2}\sqrt{B^2+C^2}} \ \frac{b}{\sqrt{\left(r^2+M^2+\frac{2M^3}{r}\right)}}
 \nonumber \\
 &&+ \frac{AB}{\sqrt{A^2-B^2-C^2}\sqrt{B^2+C^2}} \sigma \frac{r}{(r-M)}\sqrt{1-\frac{\left(b^2-M^2\right)}{r^2}+\frac{2M\left(b-M\right)^2}{r^3}}  \ ,\nonumber \\
 \nonumber \\
 U^{(\theta)}_{i,cm} &=& 0 \ .
 \label{cmmal}
\end{eqnarray}

The advantages of making transition to the center of mass frame are many fold. Firstly it allows us to implement the conservation of
energy-momentum in the collision event in a very simple way and compute the exact expressions for the conserved energy and angular momentum.
In the Boyer Lindquist coordinate system we have resport to the perturbative expansion of the conservation of energy-momentum equation in
$\epsilon$ and compute conserved energy and conserved angular momentum perturbatively to the desired order.
Secondly the differential cross-section of various particle physics processes is mentioned in the center of mass frame.

\subsection{Collision and center of mass energy}
We consider a process where two identical particles of mass $m$ collide and give rise to the two massless particles. For simplicity we
assume that the massless particles are also restricted to move onto the equatorial plane. Using the transformations given in the previous
section we can show that the total four-velocity of two colliding particles in the center of mass frame is given by
\begin{equation}
 U^{(1)}_{cm}+U^{(2)}_{cm}=\left( \sqrt{A^2-B^2-C^2}, 0,0,0 \right).
\end{equation}
Two colliding particles move in opposite directions with equal magnitude of three-velocity. The time component when multiplied by
mass of the colliding particles yields the center of mass energy of collision of two particles.
\begin{equation}
 E_{cm}=m\sqrt{A^2-B^2-C^2}.
\end{equation}
When we substitute the values of $A, B, C$ we get center of mass energy of collision to the leading order in $\epsilon$ as
\begin{equation}
 E_{cm}=m \sqrt{\frac{2(2-l_2)(2-\sqrt{2})}{\epsilon}}.
\end{equation}

\subsection{Conserved energy of collision products}

As we said earlier, we consider the case where two massless particles are produced in the collision. It follows from the transformation
that the massless particles move in the plane ($\hat{r}$-$\hat{\phi}$) in the center of mass frame if they travel in
the equatorial plane in Boyer Lindquist coordinate system. The components of four-velocity of two massless particles in the center of mass frame
\begin{eqnarray}
 U^{(\mu)}_{3,cm}= m\frac{\sqrt{A^2-B^2-C^2}}{2}~\left(1,\cos \alpha,\sin \alpha,0\right) \ ,
\label{p3}
\end{eqnarray}
and
\begin{eqnarray}
 U^{(\mu)}_{4,cm}= m\frac{\sqrt{A^2-B^2-C^2}}{2}~\left(1,-\cos \alpha,-\sin \alpha,0\right) \ ,
\label{p4}
\end{eqnarray}
where $\alpha$ is the angle between the direction along which particle-3 travels and $\hat{r}$ direction. Two particles
produced in the collision travel in opposite direction. It can be easily checked that the four-velocities written above are null and
satisfy the conservation of energy-momentum.

Expressions of the Killing vector $k=\partial_{t}$ in the center of mass frame can be written as
\begin{eqnarray}
 k^{(t)}_{cm}&=& \frac{A}{\sqrt{A^2-B^2-C^2}}\frac{(r-M)}{ \sqrt{\left(r^2+M^2+\frac{2M^3}{r}\right)}}
 +\frac{C}{\sqrt{A^2-B^2-C^2}}\frac{\frac{2M^2}{r}}{\sqrt{r^2+M^2+\frac{2M^3}{r}}} \ , \nonumber \\
  k^{(\phi)}_{cm}&=& -\frac{B}{\sqrt{B^2+C^2}} \frac{\frac{2M^2}{r}}{\sqrt{\left(r^2+M^2+\frac{2M^3}{r}\right)}} \ , \nonumber \\
 k^{(r)}_{cm}&=& -\frac{\sqrt{B^2+C^2}}{\sqrt{A^2-B^2-C^2}} \frac{(r-M)}{\sqrt{r^2+M^2+\frac{2M^3}{r}}}
 - \frac{AC}{\sqrt{A^2-B^2-C^2}}\frac{\frac{2M^2}{r}}{\sqrt{B^2+C^2}} \ , \nonumber \\
 k^{(\theta)}_{cm}&=&0 \ .
 \label{kcm}
\end{eqnarray}

We now write the expressions for the conserved energies of the particles produced in the collisions as
\begin{eqnarray}
 E_{3} &=& -\eta_{\mu\nu}P^{(\mu)}_{3,cm}k^{(\nu)}_{cm} \nonumber \\
       &=& \frac{m}{2}\frac{(r-M)A+\frac{2M^2}{r}C}{\sqrt{\left(r^2+M^2+\frac{2M^3}{r}\right)}}
       + \frac{m}{2}\frac{(r-M)\sqrt{B^2+C^2}+\frac{2M^2}{r}\frac{AC}{\sqrt{B^2+C^2}}}{\sqrt{\left(r^2+M^2+\frac{2M^3}{r}\right)}}\cos\alpha \nonumber \\
       &&+ \frac{m}{2}
        \frac{\frac{2M^2}{r}}{\sqrt{\left(r^2+M^2+\frac{2M^3}{r}\right)}}\frac{B}{\sqrt{B^2+C^2}}\sqrt{A^2-B^2-C^2}
        \sin \alpha \ ,
\label{E3}
\end{eqnarray}
and
\begin{eqnarray}
 E_{4} &=& -\eta_{\mu\nu}P^{(\mu)}_{4,cm}k^{(\nu)}_{cm} \nonumber \\
       &=& \frac{m}{2}\frac{(r-M)A+\frac{2M^2}{r}C}{\sqrt{\left(r^2+M^2+\frac{2M^3}{r}\right)}}
       - \frac{m}{2}\frac{(r-M)\sqrt{B^2+C^2}+\frac{2M^2}{r}\frac{AC}{\sqrt{B^2+C^2}}}{\sqrt{\left(r^2+M^2+\frac{2M^3}{r}\right)}}\cos\alpha \nonumber \\
       &&- \frac{m}{2} \frac{\frac{2M^2}{r}}{\sqrt{\left(r^2+M^2+\frac{2M^3}{r}\right)}}\frac{B}{\sqrt{B^2+C^2}}\sqrt{A^2-B^2-C^2} \sin \alpha \ .
\label{E4}
\end{eqnarray}

\section{White-holes as source of ultra high energy particles}

\subsection{Determining direction of motion in center of mass frame for massless particles}
The direction along which the particle moves in the center of mass frame can be determined if we know its impact parameter $b$
and based on whether or not it moves in radially outward or inward direction in the center of mass frame. Once $b$ and $\sigma$
are specified we can write down the components of velocity in the center of mass frame along $\hat{r}$ and $\hat{\phi}$ as
 \begin{eqnarray}
 U^{(r)}_{cm} &=& -\frac{\sqrt{B^2+C^2}}{\sqrt{A^2-B^2-C^2}} \ \frac{1}{(r-M)}\frac{\left(\left(r^2+M^2+\frac{2M^3}{r}\right)-\frac{2Mb}{r}\right)}{\sqrt{\left(r^2+a^2+\frac{2M^3}{r}\right)}} \nonumber \\
  &&+ \frac{AC}{\sqrt{A^2-B^2-C^2}\sqrt{B^2+C^2}} \ \frac{b}{\sqrt{\left(r^2+M^2+\frac{2M^3}{r}\right)}}
 \nonumber \\
 &&+ \frac{AB}{\sqrt{A^2-B^2-C^2}\sqrt{B^2+C^2}} \sigma \frac{r}{(r-M)}\sqrt{1-\frac{\left(b^2-M^2\right)}{r^2}+\frac{2M\left(b-M\right)^2}{r^3}}  \ ,\nonumber \\
 \nonumber \\
U^{(\phi)}_{cm} &=& -\frac{C}{\sqrt{B^2+C^2}} \ \sigma \frac{r}{(r-M)} \sqrt{1-\frac{\left(b^2-M^2\right)}{r^2}+\frac{2M\left(b-M\right)^2}{r^3}} \nonumber \\
 &&+ \frac{B}{\sqrt{B^2+C^2}} \ \frac{b}{\sqrt{\left(r^2+M^2+\frac{2M^3}{r}\right)}}  \ , \nonumber  \\ 
 \label{dir1}
 \end{eqnarray}

Given $\hat{r}$ and $\hat{\phi}$ components of velocity we can determine the direction along which massless particle moves in the center of mass frame as
 \begin{equation}
\sin \alpha = \frac{ U^{(\phi)}_{cm}}{ \sqrt{U^{(r) \ 2}_{cm}+ U^{(\phi) \ 2}_{cm}}} \ ; \  \cos \alpha = \frac{ U^{(r)}_{cm}}{ \sqrt{U^{(r) \ 2}_{cm}+ U^{(\phi) \ 2}_{cm}}}
 \label{alphamal}
\end{equation}
Once we know the value of $\alpha$ we can determine its energy.

\begin{figure}
\begin{center}
\includegraphics[width=0.7\textwidth]{wh.eps}
\caption{ The emission-cone along which massless particles must be emitted in the center of mass frame to fall into black hole, turn
back and emerge out of white hole is depicted in the figure. The emitted particle away from positive or negative $\hat{r}$
axis has divergent conserved energy.}
\label{ece}
\end{center}
\end{figure}

\subsection{Desired particles with large conserved energy}
We are interested in the massless particles produced in the collision with impact parameter in the range $M < b < 2M$ which move in the radially inward
direction. We try to obtain the emission-cone in the center of mass frame into which these particles are directed.

When we substitute
$b=2M$ and $\sigma=-1$ in Eq.\ref{dir1} we get to the leading order
\begin{equation}
 U^{(r)}_{cm}=-\frac{\sqrt{(2-l_2)}(2-\sqrt{3})}{\sqrt{2(2-\sqrt{2})}}\frac{1}{\sqrt{\epsilon}}\to -\infty ~~,~~
 U^{(\phi)}_{cm}=-1.
\end{equation}
This particle travels along the direction making angle
\begin{equation}
 \alpha=\pi +\theta ,
\end{equation}
where
\begin{equation}
 \sin\theta=\frac{\sqrt{2(2-\sqrt{2})}}{(2-\sqrt{3})}\sqrt{\epsilon}\to 0~~,~~\cos\theta \to 1.
\end{equation}
This direction almost coincides with negative $\hat{r}$ axis.

When we substitute $b=M$ and $\sigma=-1$, we get to the leading order
\begin{equation}
 U^{(r)}_{cm}=\frac{\sqrt{2-\sqrt{2}}}{\sqrt{2(2-l_2)}}\frac{1}{\sqrt{\epsilon}} \rightarrow +\infty~~,~~
 U^{(\phi)}_{cm}=\frac{l_2}{2-l_2}.
\end{equation}
$U^{(\phi)}_{cm}$ is positive or negative depending on whether $l_2$ is positive or negative in the allowed range.
This particle travels along the direction for which $\theta$ is given by
\begin{equation}
 \sin \theta \to \frac{l_2 \sqrt{2}}{\sqrt{(2-l_2)}\sqrt{(2-\sqrt{2})}}\sqrt{\epsilon} \to 0~~,~~\cos{\theta} \to -1.
\end{equation}
This direction almost coincides with positive $\hat{r}$ axis.

The cone along which particle must travel is given by part of the lower plane enclosed by the two directions
stated above. It is depicted in the Fig.\ref{ece}.

We consider the case where massless particle is emitted along the direction in the lower half of the plane with the
angle $\theta$ away from the values $0$ and $\pi$ so that both $\sin\theta$ takes a positive non-zero finite value. We compute the
value of the conserved energy which to the leading order turns out to be
\begin{equation}
 E=\frac{m}{2}\sqrt{\frac{2(2-l_2)(2-\sqrt{2})}{\epsilon}} \sin\theta \to \infty.
\end{equation}
It is clear from the expression above that the conserved energy shows divergence. Thus almost all the particles emitted in the lower
plane have divergent conserved energy which is of the order of center of mass energy. On the other hand it can be shown easily that the massless particles
emitted in the upper plane along the angle away from the values $0$ and $\pi$ admit large negative value of conserved energy.
Thus a given particle produced in the collision will admit large positive value of conserved energy with the probability slightly below
half if we assume the isotropic emission of particles in the center of mass frame. This particle would enter the event horizon and turn back.

\subsection{Emission of ultra high energy particles from white hole}
We consider the high energy collision near the event horizon of the extremal Kerr black hole in which two massless particles are produced.
If we assume the isotropic emission of massless particles in the center of mass frame any given particle will have large conserved
energy comparable to the center of mass energy with the probability half. This particle would enter the event horizon, which coincides
with the Cauchy horizon and will turn back. It emerges from the white hole event horizon into the next asymptotic region. Since the
conserved energy is preserved in this process it retains its large conserved energy. Observer at infinity perceives it as ultra-high energy
particle since the conserved energy is interpreted as the energy measured by the observer at infinity. Thus extremal Kerr white holes
would serve as the source of ultra high energy particles.

In this paper we presented a full analysis for the particles that collide outside the event horizon of extremal Kerr black hole.
Alternatively we can also consider the process where two particles collide slightly below the event horizon or below and above the
white hole event horizon. This requires one of the particle to have conserved angular momentum slightly below the critical value
considered earlier in the paper. Even these cases would lead to the production of ultra-high energy particles.

\section{Conclusion}

The origin of ultra-high energy particles such as cosmic rays and neutrinos remains an enigma and is an unsolved problem. The mechanism
that can explain the origin of high energy particles such as Fermi acceleration and unipolar inductors make use of electromagnetic forces
to energize charged particles to high energies. In this paper we made an attempt to come up with a process that is essentially
gravitational in nature rather than electromagnetic one.

We consider an extremal Kerr spacetime whose maximal extension consists of a series of infinitely many asymptotically flat regions of spacetime
joined to each other by the region which is perceived as black hole by the observers in one region and white hole by the observers
in the next asymptotic region. We try to exploit this feature to come up with a mechanism to generate high energy particles.
We consider two identical massive particles which move towards the black hole starting from rest at infinity and
collide just outside the event horizon. If the conserved angular momentum of one of the particles is fine-tuned to a critical
value and if the collision takes place at the location arbitrarily close to the event horizon, the center of mass energy of collision
diverges. We consider a process where two massless particles are produced in the collision. Assuming the isotropic emission of
collision products we show that any given particle admits divergent value of conserved energy comparable to the center of mass energy with
probability slightly below half. We show that such a particle enters the black hole event horizon which coincides with the Cauchy horizon,
turns back and emerges out of white hole event horizon into the next asymptotic region of spacetime with exactly the same value of
conserved energy. It is perceived as the ultra high energy particle by the observer at infinity. Thus the extremal white hole
appears to be the source of ultra high energy particles. In this analysis we assume that the collision takes place at the location slightly above
the black hole event horizon. However the situations where collision takes place slightly below the event horizon or slightly below and above of
white hole event horizon also give rise to ultra high energy particles, which seem to emerge out of white hole.

Similar process might also occur in the near extremal Kerr spacetime. It follows from the fact that it is possible to have collisions with large center of mass energy in the vicinity of event horizon of near-extremal black hole \cite{Harada} and the conserved energy of the particles produced in the collision is of the order of center of mass energy \cite{Patil}. Since collision takes place close to the event horizon particles with large center of mass are expected to enter event horizon and emerge out of white hole event horizon. However it would be slightly more involved since there are two small parameters in the problem, namely deviation of spin parameter from the extremality and the proximity of the collision event from the event horizon, as opposed the extremal case where the proximity of collision from the event horizon is the only small parameter making the perturbation analysis simpler. Further the global structure of near extremal black hole is significantly more complicated as compared to the extremal black hole. We shall take it up in near future. In the realistic astrophysical scenario there is a bound on how close the spin parameter could be to the extremal value \cite{Thorne}. It can possibly be violated \cite{Abm1},\cite{Abm2}. But one may not expect spin parameter to be sufficiently close to the extremal value and thus limiting the scope of the process we developed in this paper.

\section{acknowledgement}
MP would like to thank N. Dadhich for comments. MP acknowledges support from Seed Grant and Networking Fund of IIT Dharwad.
TH thanks T. Igata, Y. Koga, T. Kokubu, and K. Ogasawara for fruitful discussion. TH was supported by JSPS KAKENHI Grant Numbers JP19K03876 (TH).

\end{document}